\newcommand\cote{C{\^ o}t{\' e}}
\newcommand\jordan{Jord{\'a}n}
\newcommand\etal{{et~al.}} 
\newcommand\mM{\ifmmode(m{-}M)\else$(m{-}M)$\fi}

\newcommand\hst{{\it HST}}
\newcommand\zacs{\ifmmode z_{850}\else$z_{850}$\fi}
\newcommand\iacs{\ifmmode i_{775}\else$i_{775}$\fi}
\newcommand\gacs{\ifmmode g_{475}\else$g_{475}$\fi}
\newcommand\racs{\ifmmode r_{625}\else$r_{625}$\fi}
\newcommand\vacs{\ifmmode V_{606}\else$V_{606}$\fi}
\newcommand\gz{{\ifmmode{(g_{475}{-}z_{850})}\else$(g_{475}{-}z_{850})$\fi}}
\newcommand\gzacs{{\ifmmode{g_{475}{-}z_{850}}\else$g_{475}{-}z_{850}$\fi}}
\newcommand\riacs{{\ifmmode{r_{625}{-}i_{775}}\else$r_{625}{-}i_{775}$\fi}}
\newcommand\rzacs{{\ifmmode{r_{625}{-}z_{850}}\else$r_{625}{-}z_{850}$\fi}}
\newcommand\izacs{{\ifmmode{i_{775}{-}z_{850}}\else$i_{775}{-}z_{850}$\fi}}
\newcommand\vzacs{{\ifmmode{V_{606}{-}z_{850}}\else$V_{606}{-}z_{850}$\fi}}
\newcommand\vi{{\ifmmode{(V{-}I)}\else$(V{-}I)$\fi}}
\newcommand\Nbar{\ensuremath{\overline{N}}}
\newcommand\Lbar{\ensuremath{\overline{L}}}
\newcommand\Ltot{\ensuremath{L_{\rm tot}}}

\newcommand\Nzbar{\ensuremath{\overline{N}_{z}}}

\newcommand\zbar{\ensuremath{\overline{z}_{850}}}

\newcommand\Ibar{\ensuremath{\overline{I}}}
\newcommand\Bbar{\ensuremath{\overline{B}}}

\newcommand\mbari{\ifmmode\overline{m}_I\else$\overline{m}_I$\fi}
\newcommand\mbarI{\ifmmode\overline{m}_I\else$\overline{m}_I$\fi}
\newcommand\mbarz{\ifmmode\overline{m}_z\else$\overline{m}_z$\fi}
\newcommand\mbar{\ifmmode\overline{m}\else$\overline{m}$\fi}

\newcommand\Mbar{\ifmmode\overline{M}\else$\overline{M}$\fi}
\newcommand\lbar{\ifmmode\overline{L}\else$\overline{L}$\fi}
\newcommand\Mbarz{\ifmmode\overline{M_z}\else$\overline{M}_z$\fi}
\def\aj{AJ}
\def\apj{ApJ}
\def\apjs{ApJS}

\def\mnras{MNRAS}

\def\aap{{A\&A}}

\documentclass[
    ,final            % use final for the camera ready runs
  ,numberedheadings % uncomment this option for numbered sections
  ]
  {aipproc}

\layoutstyle{6x9}

\begin{document}

\title{Surface Brightness Fluctuations as Stellar Population
  Indicators\footnote{Invited review talk presented at the workshop
``Probing Stellar Populations out to the Distant Universe,''
September 2008, Cefal\`u, Italy.}}


\author{John P.\ Blakeslee}{
  address={Herzberg Institute of Astrophysics,
National Research Council of Canada,
Victoria, B.C.  V9E 2E7}}

\begin{abstract}
Surface Brightness Fluctuations (SBF) can provide useful information about
the unresolved stellar content of early-type galaxies and spiral bulges.  The
absolute SBF magnitude \Mbar\ in a given passband depends on the properties
of the stellar population and can be predicted by population synthesis
models.  SBF measurements in different bandpasses are sensitive to different
evolutionary stages within the galaxy stellar population.  Near-IR SBF
magnitudes are sensitive to the evolution of stars within the AGB phase,
especially the thermally pulsing AGB, while SBF in the blue and UV are
sensitive to the hot horizontal branch and post-AGB stages.  Thus, multi-band
SBF studies can constrain
important evolutionary parameters.  Empirically, SBF data at the red end of
the optical spectrum ($i$, $z$, and potentially $y$) remain excellent
distance indicators.  I~briefly review some recent work on stellar populations
using SBF, primarily from an observational point of view.
\end{abstract}

\maketitle

%%%%%%%%%%%%%%%%%%%%%%%%%%%%%%%%%%%%%%%%%%%%
%% MAINMATTER
%%%%%%%%%%%%%%%%%%%%%%%%%%%%%%%%%%%%%%%%%%%%

\section{Introduction: the SBF Method}

% what is SBF
% a little history
% what's been going on recently
% potential for stellar pops

The surface brightness fluctuations (SBF) method measures the
intrinsic pixel-to-pixel intensity variance in a galaxy image
resulting from statistical fluctuations in the numbers and
luminosities of the stars within individual pixels.  Since
the SBF signal is convolved with the point spread function, one
measures the  Fourier space amplitude of the
power spectrum on the scale of the PSF in the galaxy-subtracted \hbox{image}.
The ratio of SBF variance to galaxy surface brightness has units of flux and 
scales inversely with the square of the galaxy distance.
% ; more distant galaxies of a given stellar population appear smoother than
% nearby ones.
This ratio is usually converted to a magnitude called \mbar.  The
distance can be determined if the absolute \Mbar, which depends on both the
photometric bandpass and the stellar population, is known
from empirical or theoretical calibration.  SBF measurements in multiple
bands can provide useful distance-independent information on the stellar
content of a galaxy. 

The SBF method was first quantified by Tonry \& Schneider (1988).  The
Cefal\'u stellar populations workshop where this contribution was
presented marked an interesting anniversary, being twenty years to
the month since the publication of that seminal work.  The first major
application of the SBF method was by Tonry \etal\ (1990) for a sample of
Virgo galaxies in the $VRI$ bandpasses.  They also made a first attempt to
predict the behavior of \Mbar\ as a function of galaxy
color.  Soon afterward, Tonry (1991) presented the first fully empirical SBF
calibration, giving $\Mbar_I$ as a function of \vi.
% based on observations of Fornax cluster galaxies.
%
Following these early efforts, a large ground-based SBF survey
(Tonry \etal\ 1997, 2001) presented a redetermination of the
empirical $I$-band SBF calibration and measured distances for 300 early-type
galaxies and spiral bulges within about 40~Mpc.  For a comprehensive review
of the first decade of SBF studies, see Blakeslee, Ajhar, \& Tonry (1999).

Although the major part of SBF research has been concerned with the
measurement of extragalactic distances, peculiar velocities, and three-dimensional
structure in the local universe, recently there has been renewed 
interest in SBF as a stellar population indicator.  This is because SBF is
sensitive to the properties of the brightest stars in a galaxy in a given
bandpass, and the detailed evolution of these brightest stars is usually not
well constrained, especially for old, metal-rich stellar populations.  There
are few if any Galactic or Magellanic star clusters where such models can be
tested directly against resolved stellar systems.

There have been several recent theoretical efforts to predict SBF
magnitudes for various bandpasses and stellar populations (Liu \etal\ 2000;
Blakeslee \etal\ 2001; Mei \etal\ 2001; Cantiello
\etal\ 2003; Mouhcine \etal\ 2005; Raimondo \etal\ 2005; Marin-Franch \&
Aparicio 2006; Lee \etal\ 2009).  Cervi\~no et al.\ (2008) have recently
made a rigorous study of the theoretical underpinnings of the SBF method.
Optical and near-IR SBF measurements for Magellanic Cloud star clusters of
varying ages also provide important tests for stellar population models
(Gonz{\'a}lez et al.\ 2004; Gonz{\'a}lez-L{\'o}pezlira et al.\
2005; Raimondo et al.\ 2005).  Although there is broad agreement in
the predictions for the most common SBF bandpasses (especially $I$ band), the
agreement among different models, and between models and observations,
worsens in the near-IR and UV/blue.   We cannot be comprehensive in
the limited space of this review, and we refer the interested reader to the
original works for details.  See also the contributions by 
M.~Cantiello, R.~Gonzalez-Lopezlira, and G.~Raimondo in this volume.
Here we simply highlight a few 
results from recent SBF work related to stellar population issues.

\section{Stellar Population Effects on SBF Magnitudes}
% 1. ACS Virgo and Fornax Surveys
% 2. UV and B-band SBF

In order to work as a distance indicator, the stellar population dependence
of the SBF magnitude must be well calibrated.  This has worked best in
the $I$ and $z$ bands, where the relation between \mbar\ and galaxy color has
the smallest scatter (e.g., Blakeslee \etal\ 2001; \cote\ \etal\ 2004).
We now discuss the most recent calibrations of the SBF method and the
difficulties (and opportunities) in modeling the stellar population effects
on SBF magnitudes at various wavelengths.

%\subsubsection*{2.1~ SBF as a Distance Indicator: Some Recent Work}

\par\bigskip\centerline{\textit{2.1~ SBF as a Distance Indicator: Some Recent Work}}
\par\medskip

Most of the recent work on SBF distances has focused either on dwarf galaxies in
southern groups and clusters observed with large aperture ground-based
telescopes (e.g., Jerjen \etal\ 2004; Mieske \etal\ 2006; Dunn \& Jerjen 2006) or on early-type
galaxies observed with the \hst\ Advanced Camera for Surveys Wide Field
Channel (ACS/WFC).  SBF studies with ACS/WFC F814W bandpass (similar to $I$)
include the first optical SBF distances out to $\sim\,$100 Mpc or beyond
(Biscardi \etal\ 2008; Blakeslee \etal, in preparation), a measurement of the
distance of the peculiar gas-rich lenticular galaxy NGC\,1533 in the Dorado group (Barber Degraaff
\etal\ 2007), and studies of SBF gradients in galaxies (Cantiello \etal\
2005, 2007b; discussed below).  

The most extensive SBF work with the ACS to date has been in the F850LP band
(\zacs) as part of the ACS Virgo and Fornax cluster surveys (\cote\ \etal\
2004; \jordan\ \etal\ 2007).  Mei \etal\ (2005, 2007) present the SBF
analysis, calibration, distance measurements for 90 early-type galaxies in
the ACS Virgo survey, and Blakeslee \etal\ (2009) present the same for 43
galaxies in the ACS Fornax survey.  The latter study also presents a
recalibration based on the combined Virgo and Fornax samples.

\begin{figure}
  \includegraphics[height=.3\textheight]{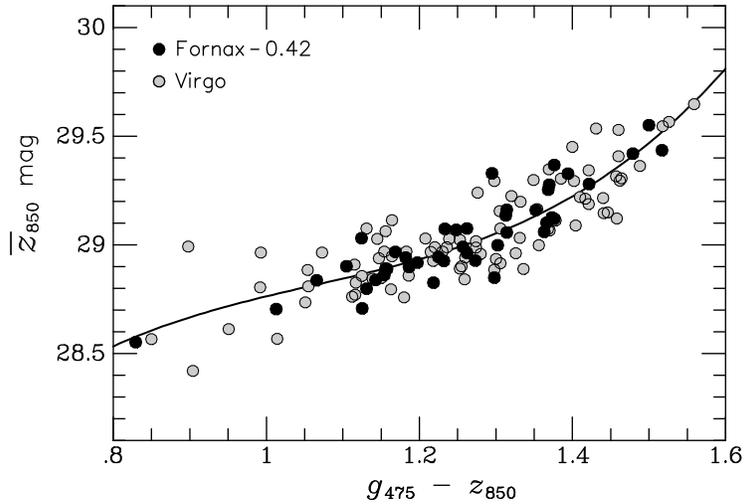}
\caption{Combined SBF \zbar\ vs \gz\ color calibration for our ACS Virgo and Fornax
  cluster survey galaxies (from Blakeslee \etal\ 2009).    SBF magnitudes for
the Fornax galaxies are shifted brighter by the best-fit relative distance
modulus of  $0.42\pm0.03$~mag.  The curve
  represents the best-fit cubic polynomial calibration for the dependence of
  \zbar\ on \gz.
\label{fig:finalcal}}
\end{figure}

Figure~\ref{fig:finalcal} shows the resulting \zacs-band SBF
calibration from Blakeslee \etal\ (2009).  The dependence of \zbar\ on \gz\
is nonlinear and can be well described by a cubic polynomial.  The fit yields
a highly precise relative distance modulus of $0.42\pm0.03$ mag (including
systematic uncertainties) for Fornax with respect to Virgo.  The intrinsic
scatter in the method is estimated empirically to be just 0.06~mag for
galaxies with $\gz>1$, but increases at bluer colors (the compactness of
Fornax makes it possible to measure this small intrinsic scatter).  
Mei \etal\ (2005) showed that the general behavior (with a zero-point shift)
of SBF in \zacs\ was reasonably approximated by predictions from Bruzual \& Charlot (2003)
simple stellar population (SSP) models of varying age and metallicity.
However, it is not possible to predict the scatter in the SBF relations from SSP
models.  Chemo-evolutionary or semi-analytic models are required for this,
although some estimates can be made from simple monte carlo simulations
(Blakeslee \etal\ 2001).  A fully successful model should predict the very
tight \zbar\ vs \gz\ relation with increased scatter at the blue end.

\begin{figure}
  \includegraphics[height=.3\textheight]{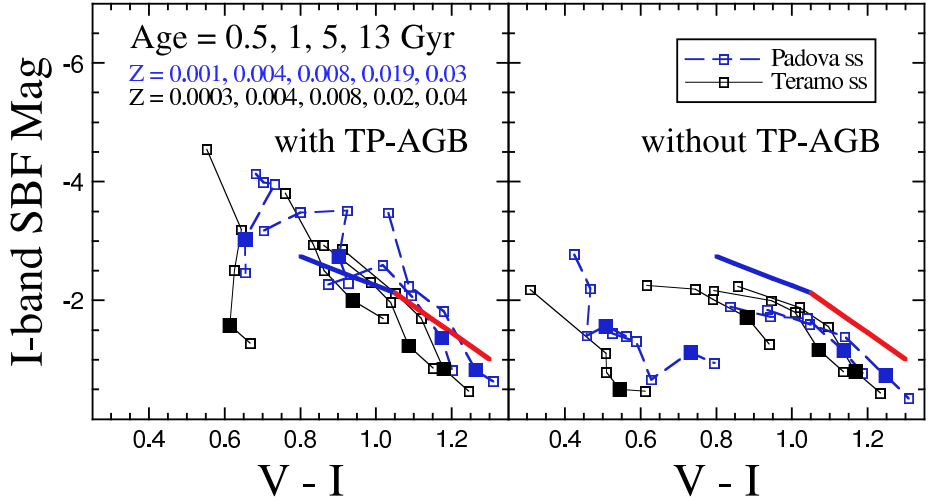}
\caption{Comparison of Padova and Teramo/BaSTI stellar population model 
SBF predictions in the $I$ band, with (left) and without (right) the thermally pulsing
AGB (TP-AGB) stage, adapted from Lee \etal\ (2009).
At each of four different ages, models with five
different metallicities are connected by the thin lines. 
For reference, the large solid squares mark the solar metallicity
models of varying age. The empirical $I$-band SBF relations are shown by the
bent thick lines (see Lee \etal\ 2009 for details).
\label{fig:padterI}}
\end{figure}

At present, significant uncertainties remain in SBF model
predictions even in well-studied bandpasses.  For instance,
Figure~\ref{fig:padterI} compares SBF predictions in the $I$ band from Padova
and Teramo/BaSTI SSP models (see Lee \etal\ 2009 for full details).  The
predictions are given both with and without the contributions from the
thermally pulsing AGB (TP-AGB) stage.  Two things are worth noting: there are
large differences between the two sets of model predictions (left panel), and
the highly uncertain TP-AGB stage has a very large effect (compare left
and right panels).  The agreement between the sets of models is better when
this stage is omitted, but of course it is needed to match the observations
(represented by the thick lines).

%\subsubsection*{2.2~ SBF in the UV/Blue: Hot Horizontal Branch and the Post-AGB}

\par\bigskip\centerline{\textit{2.2~ Multi-band SBF: Hot Horizontal Branches,
TP-AGB, Post-AGB, etc.}}\par\medskip

Worthey (1993b) showed theoretically that SBF measurements in the UV could provide
quantitative information on the presence of extended hot horizontal branch
stars in elliptical galaxies and other unresolved stellar systems.  According
to the models used by Worthey, the presence of a hot horizontal branch could
make the $U$-band SBF magnitude brighter by 3~mag in a metal-rich early-type
galaxy.   The effect at shorter wavelengths, such as the F284W band of
\hst/WFPC2, was even larger.  However, the SBF signal is very faint at these
wavelengths, and to date there have been no practical demonstrations of the
usefulness of $U$-band SBF for this purpose.  With the coming of large 
ground-based telescopes equipped with blue sensitive detectors (and hopefully
the UVIS channel of WFC3) this may soon change.  

Although the ACS/WFC has no sensitivity shortward of 4000\,\AA, this
instrument (before its sad demise) afforded the first samples of reliable
$B$-band (F435W) SBF measurements beyond the Local Group (Cantiello \etal\
2007b), which have proven to have interesting implications for stellar
population synthesis.  Cantiello \etal\ (2007b) demonstrated the sensitivity
of SBF \Bbar$-$\Ibar\ colors to the rate of post-AGB evolution.  The
observational data disagreed with standard model predictions, but Cantiello
\etal\ showed that the models could be brought into agreement with the data
by adjusting the post-AGB evolution to agree with that inferred for resolved
stars in~M32.  The hot post-AGB stars have a proportionately larger effect on
\Bbar\ than on \Ibar.  

However, this solution was not unique, since an
unmodeled hot horizontal branch or other missing components could also 
cause the observed discrepancy.  Additional SBF measurements in the UV
(to constrain further the properties of the hot component)
and near-IR (sensitive to the post-AGB star progenitors) could fully resolve this
issue.  Multi-band optical/IR
SBF data are thus 
relevant to such problems as AGB evolution and the UV excess in elliptical
galaxies (e.g., Buzzoni \& Gonzalez-Lopezlira 2008).

%% \subsubsection*{2.3~ SBF in the Near-IR and the Thermally Pulsing AGB}
%\par\bigskip\centerline{\textit{2.3~ SBF in the Near-IR and the Thermally Pulsing AGB}}
%\par\medskip

\begin{figure}
  \includegraphics[height=.3\textheight]{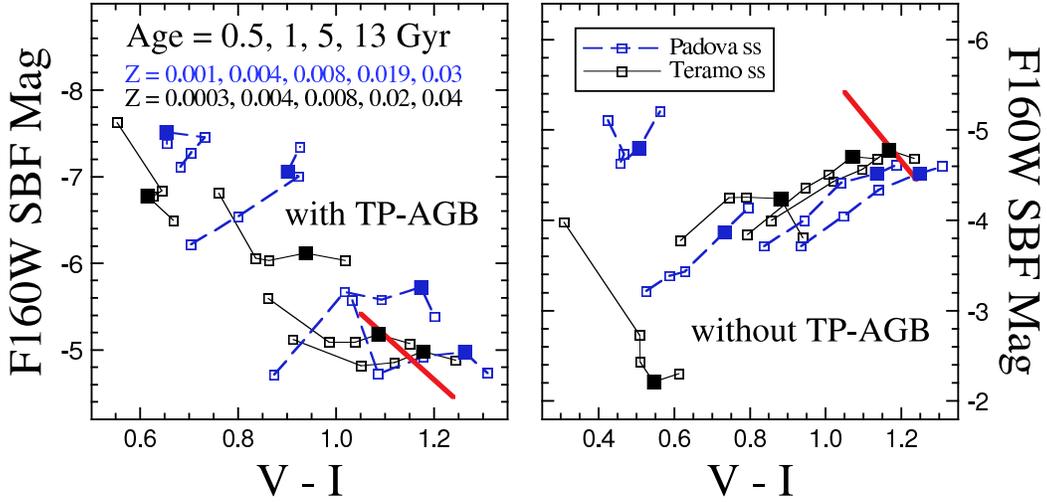}
\caption{Comparison of Padova and Teramo/BaSTI stellar population model 
SBF predictions, similar to Figure~\ref{fig:padterI}, but here shown
for the \hst\ NICMOS F160W band.  The empirical relation from Jensen \etal\ (2003) is 
shown as a thick solid line.  The importance of the TP-AGB phase is
considerably greater in the near-IR.  The effect of these stars in fact reverses
the sense of the dependence of the SBF magnitude on color,
so that the near-IR SBF brightens
as observed at bluer colors, instead of becoming fainter as it would without
the TP-AGB.  See Lee \etal\ (2009) for details.
\label{fig:padter160}}
\end{figure}

% Large uncertanties remain in the modeling of the AGB phase, particularly the
% evolution of TP-AGB stars. 
%
Blakeslee \etal\ (2001) found that the treatment of the AGB phase was
responsible for some large discrepancies in the near-IR SBF predictions of
different models; see also the discussion by Liu \etal\ (2000).
Figure~\ref{fig:padter160}, adapted from Lee \etal\ (2009) shows a dramatic
demonstration of the effects of the TP-AGB phase on near-IR SBF (in the \hst\
F160W bandpass).  For both the Padova and Teramo/BaSTI sets of models, the
inclusion of the TP-AGB phase actually reverses the sense of the dependence
of the SBF magnitude on integrated color (at least for age $\ge 1$~Gyr).

The original expectations from models were that
near-IR SBF magnitudes would brighten at redder
colors (e.g., Worthey 1993a).  It was a surprise when the reverse was found
observationally (Jensen \etal\ 1998, 2003).  The modeling of the AGB
phase was likely much of the reason for the erroneous
prediction.  For the most part, the inclusion of the AGB in population
synthesis relies on imperfectly constrained empirical prescriptions, rather
than true evolutionary models.  It is evident from the figures and 
comparisons with observations that nature has ordained a 
better behaved AGB phase 
than those included in the population models.  The relations between
SBF magnitude and integrated color are tighter than might be expected from
current models.  \hbox{Multi-band} SBF data can provide powerful constraints to aid
in the development of true AGB evolutionary codes for inclusion in the next
generation of stellar population models.

\section{SBF and Stellar Population Gradients}

Until now we have discussed mean colors and SBF magnitudes, but these
quantities vary within galaxies, most notably because of radial gradients in
stellar population properties.  SBF gradients were known from ground-based
data for a few galaxies (Tonry 1991; Sodemann \& Thomsen 1995, 1996; Tonry
\etal\ 2001), but until recently there were no systematic studies.
NGC\,205 was the first galaxy where an SBF gradient was noted.  It is 
unusual in that its center is blue and it becomes redder outward, the reverse
of typical gradients.  This is because the gradient is due to relatively recent central star
formation, and therefore an age effect.  The slope of the dependence of
\mbarI\ on \vi\ is also shallower than that of the mean dependence for
galaxies.

\begin{figure}
  \includegraphics[height=.3\textheight]{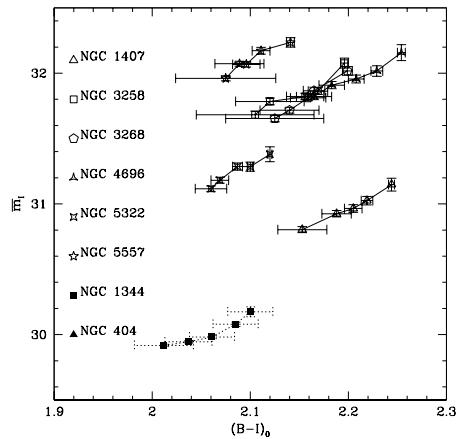}
\caption{$(B{-}I)$ colors and $I$-band SBF measurements in a series of concentric
annuli are shown for seven early-type galaxies studied by Cantiello \etal\
(2005) with the ACS/WFC.  In all of these galaxies, the color gets bluer and
\mbari\ brighter with increasing radial distance from the galaxy center.  The
vertical shifts are due to the different galaxy distances (see Cantiello
\etal\ 2005, 2007a for details).
\label{fig:gradients}}
\end{figure}

Cantiello \etal\ (2005, 2007a) have used the ACS/WFC to make the first
studies of SBF gradients in significant samples of galaxies (see
Figure~\ref{fig:gradients}).  They show that the size of the internal SBF vs
color slope for multiple annuli within a galaxy can reveal whether the
gradients are due to age or metallicity.  In most cases, they are metallicity
gradients, but in some cases an age gradient appears equally or more
important.  The galaxies with age gradients tend to be bluer and show some
evidence for past mergers (e.g., NGC\,1344).   At these optical wavelengths, age
and metallicity variations cause qualitatively similar SBF vs color relations, though
with different slopes.  It would be extremely interesting to study
internal SBF gradients in the near-IR where pure age or metallicity
variations would translate to very different dependences of SBF magnitude on color.
With improvements in near-IR instruments, this is now becoming possible.

\begin{figure}
  \includegraphics[height=.27\textheight]{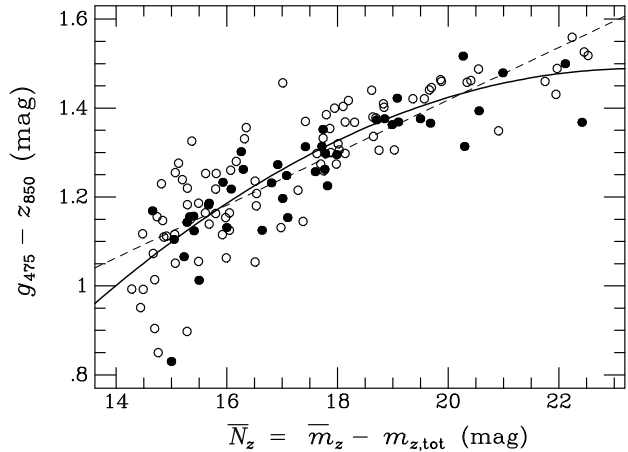}
\caption{Galaxy \gz\ color is plotted as a function of the ``fluctuation count''
$\Nzbar$, the difference between SBF magnitude \zbar\ and total $z$ magnitude
of the galaxy, for galaxies in Virgo (open circles) Fornax (filled circles).
Note that \Nzbar, like \gz, is distance-independent and can therefore be used
for an alternative calibration of the SBF method (see~Blakeslee \etal\ 2009).
Unlike color, \Nbar\ is also independent of Galactic extinction.  The dashed
and solid lines show the best-fit linear and quadratic relations,
respectively.  \Nbar\ is closely related to the stellar mass of the galaxy,
and the correlation with color results from the mass-metallicity scaling
relation; the scatter increases at blue colors and low luminosities.%
\vspace{-9pt}
\label{fig:gzNbar}}
\end{figure}

\section{SBF and Total Stellar Mass: \Nbar}
\vspace{-2pt}
The distance-independent ``fluctuation~count'' \Nbar\ is defined as 
$ \Nbar \,=\, \mbar - m_{\rm tot} \,=\,
    \hbox{+}2.5\,\log[L_{\rm tot}\,/\, \lbar\,]$ 
(Tonry \etal\ 2001), where $\Ltot$ is the galaxy total luminosity and
\Lbar\ is the mean ``fluctuation luminosity'' corresponding to  magnitude 
\Mbar, which depends on the stellar population.  The ratio $\Ltot/\Lbar$ 
is a distance-independent galaxy luminosity in units of \Lbar\ and scales
with the number of stars in the galaxy (and thus with stellar mass).
Figure~\ref{fig:gzNbar} shows the correlation of \Nbar\ with color from
Blakeslee \etal\ (2009), a consequence of the mass-metallicity relation. 
\Nbar~also shows a good correlation with velocity dispersion $\sigma$, tighter 
than even the Mg$_2$-$\sigma$ relation (Blakeslee \etal\ 2002).

The correlation with color means that \Nbar\ 
can also be used to calibrate SBF distances.  Depending on the color baseline and
properties of the data sample, the SBF calibration based on \Nbar\ may show
less scatter than that from color (Blakeslee \etal\ 2002, 2009).
However, the calibration is then no longer based purely on stellar population
properties, but involves a galaxy mass scaling relation.
Thus, luminous blue galaxies and small red galaxies deviate from the
calibration, and there may be systematic environmental effects similar those 
of the fundamental plane.  For these reasons, 
the preferred calibration uses galaxy color, but the behavior of \Nbar\ itself 
would be interesting to examine with semi-analytic galaxy modeling.

\vspace{-5pt}
\section{Outlook}
\vspace{-2pt}
Multi-band SBF studies are a unique and quite promising way of constraining stellar
population properties, especially when combined with additional information.  SBF
data could help enormously in understanding TP-AGB evolution and hot stellar
components in elliptical galaxies.  Although there has been much theoretical
modeling in this regard, the observational studies to date have been limited.  The
current lack of an efficient large-format space-based camera is a big blow, but
currently available large-format near-IR imagers and the coming generation of
multi-band digital sky surveys provide interesting prospects for new SBF
studies. ~We have only scratched the surface of this field.

% improvements in ground-based facilities

%%%%%%%%%%%%%%%%%%%%%%%%%%%%%%%%%%%%%%%%%%%%%%%%
%% BACKMATTER
%%%%%%%%%%%%%%%%%%%%%%%%%%%%%%%%%%%%%%%%%%%%%%%%
%\bigskip
% \begin{theacknowledgments}
\vspace{0.67cm}
\noindent
I wish to thank the conference organizers for inviting me to talk on
this subject.  I am indebted to my collaborators on the ACS
Virgo and Fornax cluster surveys for their many contributions to this work,
my colleagues at Teramo Observatory for their hospitality
following the conference, and Hyun-chul Lee for providing
figures prior to publication.
% \end{theacknowledgments}

\end{document}